\documentclass[12pt]{article}
\usepackage{epsfig}
\usepackage{latexsym}
\begin{document}
\begin{center}

{ \bf ON THE FEIGEL EFFECT: EXTRACTION OF MOMENTUM FROM VACUUM?}

\vspace{1cm}

Ole Jakob Birkeland and  Iver Brevik\footnote{E-mail:
iver.h.brevik@ntnu.no (corresponding author).}

\bigskip

Department of Energy and Process Engineering, Norwegian University
of Science and Technology, N-7491 Trondheim, Norway

\bigskip

\bigskip

\end{center}

\begin{abstract}

The Green-function formalism for the electromagnetic field in a
magnetoelectric (ME) medium is constructed, as a generalization of
conventional Casimir theory. Zero temperature is assumed. It is
shown how the formalism predicts electromagnetic momentum to be
extracted from the vacuum field, just analogous to how energy is
extracted in the Casimir case. The possibility of extracting
momentum from vacuum was discussed recently by Feigel [Phys. Rev.
Lett. {\bf 92}, 020404 (2004)]. By contrast to Feigel's approach,
we assume that the ME coupling occurs naturally, rather than being
 produced by external strong fields. We also find the same effect
qualitatively via another route, by considering one single
electromagnetic mode.

\end{abstract}

PACS numbers: 03.50.De, 12.20.-m, 42.50.Nn, 42.50.Vk

\bigskip

\section{Introduction}

Consider a magnetoelectric (ME) medium whose constitutive
relations can be written on compact form as
\begin{equation}
{\bf D}=\varepsilon_0 \varepsilon \cdot {\bf E}+\frac{1}{\mu_0
c}\,\chi \cdot {\bf B}, \label{1}
\end{equation}
\begin{equation}
{\bf H}=-\frac{1}{\mu_0 c}\,\chi^T \cdot {\bf E}+\frac{1}{\mu_0}\,
\mu^{-1} \cdot {\bf B}. \label{2}
\end{equation}
Here $\chi$ is the ME coupling parameter, assumed in general to be
a pseudotensor, with $(\chi^T)_{ik} \equiv \chi_{ki}$. We employ
SI units, so that the relation $\varepsilon_0 \mu_0=1/c^2$ refers
to a vacuum, and we let the permittivity tensor $\varepsilon_{ik}$
and permeability tensor $\mu_{ik}$ be nondimensional (i.e.,
relative), so that the relations $D_i=\varepsilon_0
\varepsilon_{ik} E_k$ and $B_i=\mu_0 \mu_{ik} H_k$ apply in the
non-chiral case when $\chi=0$. We shall take all material
quantities $\varepsilon_{ik}, \mu_{ik}, \chi_{ik}$ to be real and
frequency independent. The tensors $\varepsilon_{ik}$ and
$\mu_{ik}$ are symmetric; this being a general electrodynamic
property following from the symmetry of the kinetic coefficients
\cite{landau84}. No such symmetry condition exists for $\chi$,
however. In some materials $\chi_{ik}$ is symmetric,
$\chi_{ik}=g\delta_{ik}$ with $g$ a pseudoscalar function, or,
what is of more interest in the present context, $\chi_{ik}$ can
be antisymmetric. An anisotropic crystal is called biaxial if the
diagonal permittivity components $\varepsilon_x \neq \varepsilon_y
\neq \varepsilon_z$ along the principal axes, and is called
uniaxial if $\varepsilon_x = \varepsilon_y \neq \varepsilon_z$.

In the following we will focus attention on the situation where
the  anisotropy in $\chi_{ik}$ occurs {\it naturally}. Cases where
the anisotropy is created artificially, by means of strong
electric and magnetic fields perpendicular to the direction of
light propagation (cf., for instance, Ref.~\cite{roth02}), are for
the most part  outside the scope of the present paper.

The macroscopic theory of ME media has been known for a long time.
The reader may consult the book of O'Dell, for instance
\cite{dell70}, as well as classic papers \cite{fuchs65,dell65}. A
recent review is given by Fiebig \cite{fiebig05}; other relatively
recent papers are
Refs.~\cite{rikken00,figotin01,roth02a,tiggelen07}. As explained
in the Fiebig paper, two major sources for "large" ME effects can
be identified: (i) In composite materials the ME effect is
generated as a product property of a magnetostrictive and a
piezoelectric compound. A linear ME polarization is induced by a
weak ac magnetic field oscillating in the presence of a strong dc
bias field. (ii) In multiferroics the internal magnetic and/or
electric fields are enhanced by multiple long-range ordering. The
ME effect can be strong enough to trigger magnetic or electrical
phase transitions.

The recent paper of Feigel \cite{feigel04} - cf. also the comments
 \cite{schutzhold04,feigel04a,tiggelen04a,feigel04b} - sharpened the
 interest in this special kind of materials. The main
idea of this paper was to suggest a new quantum mechanical effect,
namely the extraction of material {\it momentum} from the
electromagnetic vacuum oscillations. The suggested effect is thus
analogous to the well known Casimir effect \cite{casimir48}, in
which case it is an energy, not a momentum, that is extracted from
the vacuum field. The Feigel effect thus belongs to a very active
area in modern physics. Its main theme is  the observability and
the interpretation of vacuum-induced phenomena in macroscopic
media.
 The effect has moreover a bearing on the famous Abraham-Minkowski
energy-momentum problem in dielectric matter
\cite{minkowski10,abraham10}.

And this brings us to the main topic of the present paper, which
is to  investigate how the Green function approach, frequently
used in Casimir-related problems, can be applied to a ME medium.
To our knowledge, such a general approach has not been developed
before. We follow the same basic field theoretical method as in
the recent paper of Ellingsen and Brevik \cite{ellingsen07},
dealing with the Casimir effect. We will show that, even in the
presence of the complexity in formalism caused by the ME effect,
the theory leads to a right/left asymmetry in a medium-filled
cavity enclosed within conducting walls placed at positions $z=0$
and $z=a$, and thus permits the extraction of momentum from the
vacuum field, in principle. Our field theoretical formalism thus
supports earlier results that were based upon consideration of
particular  modes only. We will also have the opportunity to
comment occasionally on some of the  papers that followed the
Feigel paper \cite{tiggelen04,rikken05,tiggelen06}.

In sections 2 and 3 we establish the governing equation for the
Green functions, relate this to the two-point functions for the
electromagnetic fields, and give explicit solutions in the
presence of the two conducting plates. In Sect.~4 we digress to
consider the momentum conservation equation for a ME medium, and
show how the right/left momentum asymmetry occurs for one single
mode. In Sect.~5  we return to the Green-function approach, and
show how the momentum asymmetry occurs also for the vacuum field,
when summing over all modes propagating in the $\pm x$ directions.

We thus discuss the momentum asymmetry via two different
approaches. A more detailed overview of the outline of the paper
is given in Sect.~6.

Readers interested in recent reviews on the Casimir effect may
consult
Refs.~\cite{lamoreaux05,milton01,bordag01,milton04,nesterenko04,capasso07}.
Much information can also be found in the recent special issues of
J. Phys. A \cite{jphysa06} and New. J. Phys. \cite{newjphys06}.

We emphasize that the formalism below is constructed from the same
main standpoint as in conventional Casimir theory: we calculate
the change in field momentum caused by the {\it geometric
boundaries, i.e., the plates}. The undisturbed system with respect
to which we regularize Green-function expressions is an infinite
medium (without plates), made up of the same material. It is thus
clear that in the limit when the separation between the plates
goes to infinity, the effect that we calculate has to go to zero.

\section{Governing equations for Green's function}

In this section we will establish the governing equations for the
retarded Green function in the chiral medium. When this function
is known, one can find the electromagnetic two-point functions and
thus construct expressions for energy and momentum in the field.
From now on, we assume the material to be isotropic, so that
$\varepsilon_{ik}=\varepsilon \delta_{ik},\, \mu_{ik}=\mu
\delta_{ik}$. Important in our context  is that the coupling
parameter $\chi_{ik}$ will still be permitted to be anisotropic.
As already mentioned we take all material parameters $\varepsilon,
\mu, \chi_{ik}$ to be real and frequency independent. They will
moreover be assumed to be independent of the spatial coordinates.
Our medium is thus assumed to be spatially homogeneous but chiral.
(If the anisotropy of $\chi_{ik}$ is created artificially, by
means of strong crossed electric and magnetic fields, the
anisotropy  property of $\chi_{ik}$ holds of course only in the
constant field region between the condenser plates.)

Let us first invert the constitutive relations (\ref{1}) and
(\ref{2}) to get
\begin{equation}
{\bf E}=\frac{1}{\varepsilon \varepsilon_0}\left({\bf
D}-\frac{\mu}{c}\,{\bf \chi} \cdot {\bf H} \right), \label{3}
\end{equation}
\begin{equation}
{\bf B}=\mu \mu_0 \left({\bf H}+\frac{c}{\varepsilon}\,\chi^T
\cdot {\bf D}\right). \label{4}
\end{equation}
These expressions hold when the EM effect is small, $|\chi_{ik}|
\ll 1$, what in practice always is the case. Terms of order
$\chi^2$ are neglected.

Consider now Maxwell's equations in conventional form
\begin{equation}
{\bf \nabla \cdot D}=\rho, \quad {\bf \nabla \cdot B}=0, \label{5}
\end{equation}
\begin{equation}
{\bf \nabla \times E}=-{\bf \dot B}, \quad {\bf \nabla \times
H=J+\dot D}, \label{6}
\end{equation}
and take the curl of the first member of (\ref{6}). Observing
Eq.~(\ref{4}) we then get, when neglecting terms of order $\chi^2$
throughout, the following coupled vector equation for the basic
fields $\bf E$ and $\bf B$
\begin{equation}
{\bf \nabla \times \nabla \times E}+\frac{\varepsilon
\mu}{c^2}\,{\bf \ddot E}+\frac{\mu}{c}\chi \cdot {\bf \ddot
B}+\frac{\mu}{c}\,{\bf \nabla \times (}\chi^T {\bf \cdot \dot
E)}=-\mu \mu_0 {\bf \dot J}. \label{7}
\end{equation}
If $\chi=0$, the coupling between the fields is absent. On
component form the equation can be written
\begin{equation}
\nabla^2 E_i- \partial_i\,({\bf \nabla \cdot E})-\frac{\varepsilon
\mu}{c^2}\ddot{E}_i-\frac{\mu}{c}\chi_{ik}\ddot{B}_k
-\frac{\mu}{c}\chi_{lk}  \, curl_{ik} \dot{E}_l =\mu
\mu_0\dot{J}_i. \label{8}
\end{equation}
We have here defined $curl_{ik} \equiv \epsilon_{ijk} \partial_j$,
where $\epsilon_{ijk}$ is the antisymmetric Levi-Civita symbol
with $\epsilon_{ijk}=1$.

In Eq.~(\ref{8}), the magnetic field $B_k$ can actually be
replaced by electric field components in view of one of Maxwell's
equations, $\dot{B}_k=-curl_{kl} E_l$. We obtain
\begin{equation}
\left[
\delta_{il}\nabla^2-\partial_i\partial_l-\delta_{il}\frac{\varepsilon
\mu}{c^2}\,\partial_t^2+\frac{\mu}{c}\chi_{ik}\,curl_{kl}\,\partial_t-\frac{\mu}{c}\chi_{lk}\,curl_{ik}\,\partial_t
\right]E_l=\mu\mu_0\dot{J}_i, \label{9}
\end{equation}
with $\partial_t=\partial/\partial t$.

We now turn to the Green-function approach. According to the
source theory of Schwinger {\it et al.} (see, for instance,
Refs.~\cite{milton01} or \cite{schwinger78}), we make the
correspondence ${\bf J} \rightarrow \dot{\bf P}$.\, $\rho
\rightarrow -{\bf \nabla \cdot P}$. We introduce a dyad ${\bf
\Gamma} (x,x')$ such that
\begin{equation}
{\bf E}(x)=\frac{1}{\varepsilon_0}\int d^4 x'\, {\bf
\Gamma}(x,x')\cdot {\bf P}(x'), \label{10}
\end{equation}
where $x=({\bf r}, t)$. Due to causality, $t'$ is only integrated
over the region $t' \leq t$. The dyad $\bf \Gamma$ is the retarded
Green function; also called the generalized susceptibility. We
take the Fourier transform of $\bf \Gamma$,
\begin{equation}
{\bf \Gamma}(x,x')=\int_{-\infty}^\infty
\frac{d\omega}{2\pi}\,e^{-i\omega \tau} \,{\bf
\Gamma(r,r'},\omega), \quad \tau= t-t', \label{11}
\end{equation}
exploiting the stationarity of the system. We transform also the
electric field,
\begin{equation}
{\bf E}(x)=\int_{-\infty}^\infty \frac{d\omega}{2\pi}\,
e^{-i\omega t}\, {\bf E(r}, \omega), \label{12}
\end{equation}
with a similar expression for ${\bf P}(x)$. The governing equation
for the Green function then becomes
\[ {\bf \nabla \times \nabla \times
\Gamma(r,r'},\omega)-\frac{\varepsilon \mu \omega^2}{c^2}{\bf
\Gamma(r,r'},\omega) +\frac{i\mu \omega}{c}\chi \cdot [{\bf \nabla
\times \Gamma(r,r'},\omega)]\]
\begin{equation}
-\frac{i\mu \omega}{c}{\bf \nabla \times [\chi}^T{\bf  \cdot
\Gamma(r,r'},\omega)]=\frac{\mu \omega^2}{c^2}\delta( {\bf
r-r'}){\bf 1}, \label{13}
\end{equation}
or, on component form,
\[ \Big[
\partial_i \partial_j-\delta_{ij}\nabla^2-\frac{\varepsilon \mu \omega^2}{c^2}\delta_{ij}+\frac{i\mu\omega}{c}\,\chi_{il}\,curl_{lj}
 \]
 \begin{equation}
 -\frac{i\mu\omega}{c}\chi_{jl}\,curl_{il} \Big] \Gamma_{jk}({\bf
 r,r'},\omega)=\frac{\mu\omega^2}{c^2}\,\delta({\bf
 r-r'})\delta_{ik}. \label{14}
 \end{equation}
 If $\chi_{ik}=0$ and $\mu=1$, this equation reduces to
 Eq.~(75.16) in Ref.~\cite{lifshitz80} (their symbol $D_{ik}$ is the
 same as our $ -\hbar c^2\Gamma_{ik}/\omega^2$).

 If $\Gamma_{ik}({\bf r,r'},\omega)$ is known, we can make use of
 the fluctuation-dissipation theorem (which has a meaning both classically and quantum mechanically;
 cf. Refs.~\cite{lifshitz80,landau85}), to calculate the two-point
 functions:
 \begin{equation}
 i\langle E_i({\bf r})E_k({\bf
 r'})\rangle_\omega=\frac{\hbar}{\varepsilon_0}{\rm Im}
 \{\Gamma_{ik}({\bf r,r'},\omega)\}, \label{15}
 \end{equation}
 \begin{equation}
 i\langle B_i({\bf r}) B_k({\bf
 r'})\rangle_\omega=\frac{\hbar}{\varepsilon_0}\frac{1}{\omega^2}\,curl_{ij}
 curl'_{kl} \,{\rm Im} \{\Gamma_{jl}({\bf r,r'},\omega)\}, \label{16}
 \end{equation}
 \begin{equation}
 \langle E_i({\bf r}) B_k({\bf
 r'})\rangle_\omega=\frac{\hbar}{\varepsilon_0}
 \frac{1}{\omega}curl'_{kl}\,{\rm Im}\{\Gamma_{il}({\bf
 r,r'},\omega)\}. \label{17}
 \end{equation}
 Here $curl'_{ik}=\epsilon_{ijk}\partial'_j$, where $\partial'_j$ is
 the derivative with respect to  component $j$ of $\bf r'$. The
 expressions (\ref{15})-(\ref{17}) refer to zero temperature; a
 factor sgn($\omega$) is omitted throughout. The spectral
 correlation tensor $\langle E_i({\rm r})E_k({\rm r'})\rangle_\omega$
 is defined according to
 \begin{equation}
\langle E_i(x)E_k(x')\rangle =\int_{-\infty}^\infty
\frac{d\omega}{2\pi}\,e^{-i\omega \tau}\langle E_i({\bf
r})E_k({\bf r'})\rangle_\omega. \label{18}
\end{equation}
[Note the meaning of the formalism here: the spectral correlation
tensor is related to the Fourier transform $ \langle  E_i({\bf
r},\omega)E_k({\bf r'},\omega')\rangle$ of the two-point function
$\langle E_i(x)E_k(x')\rangle$ via
\begin{equation}
\langle  E_i({\bf r},\omega)E_k({\bf r'},\omega')\rangle =2\pi
 \langle E_i({\bf r})E_k({\bf r'})\rangle_\omega \delta
(\omega+\omega'); \label{18a}
\end{equation}
cf. Eq.~(122.12) in Ref.~\cite{landau85} or also Appendix B in
Ref.~\cite{brevik99}.]

Before going on to solve these equations, we will specify the
geometry to be assumed in the rest of this paper.

\section{Specification of the geometry. Solutions for the Green
functions}

Let us assume the same setup as in conventional Casimir theory,
namely two perfectly conducting parallel plates separated by a gap
$a$. The geometry is sketched in Fig.~1.

\begin{figure}
\begin{center}
\includegraphics[width=4in]{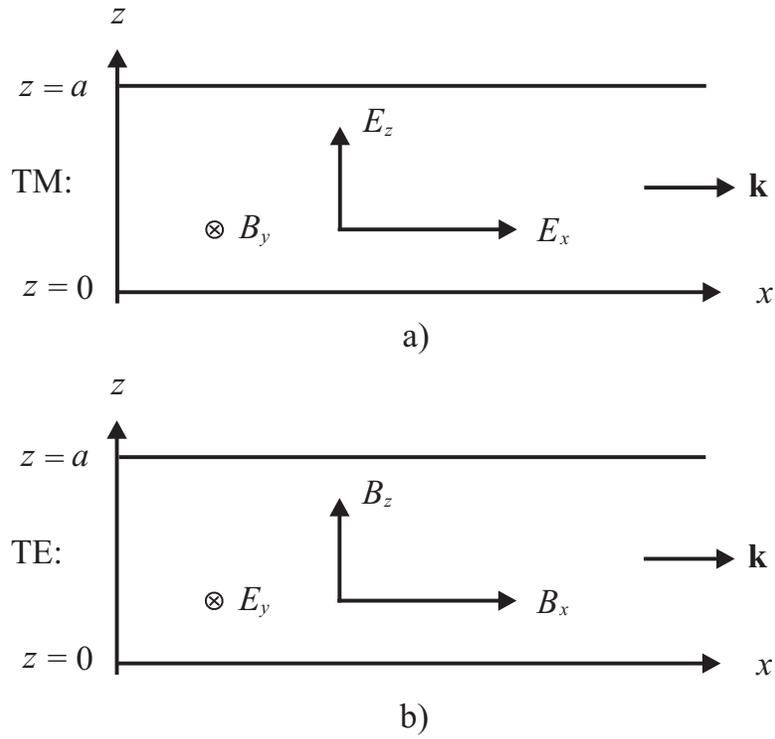}
\caption{Sketch of the geometry. The TM and TE modes are shown.
The tensor $\chi_{ik}$, constant everywhere in the fluid, is given
by Eq.~(\ref{19}). The wave vector $\bf k$ is directed along the
$x$ axis.}
\end{center}
\end{figure}

 As mentioned earlier, we will mainly be considering the case
where the ME effect occurs naturally. We assume accordingly that
$\chi_{ik}$ is given initially and is constant everywhere in the
fluid, on the inside as well as on the outside of the plates.
Because of the translational invariance in the $x$ and $y$
directions we can transform the Green function once more to obtain
\begin{equation}
{\bf \Gamma (r,r'},\omega)=\int \frac{d^2k}{(2\pi)^2}\, e^{i{\bf k
\cdot (r-r')}}\,{\bf g}(z,z',{\bf k},\omega). \label{19}
\end{equation}
We also transform the delta function:
\begin{equation}
\delta ({\bf r-r'})=\int \frac{d^2k}{(2\pi)^2}\, e^{i{\bf k \cdot
(r-r')}}\,\delta(z-z'), \label{20}
\end{equation}
and assume that $\chi_{ik}$ has the following form:
\begin{equation}
\chi_{ik}=
\left( \begin{array}{ccc} 0 & 0 & 0 \\
0 & 0 & \chi_{yz} \\
0 & \chi_{zy} & 0
\end{array}
\right) \label{21}
\end{equation}
Our coordinate system is thus henceforth fixed, relative to the
material. We focus attention on only one particular wave number
$\bf k$ in the following, namely ${\bf k}=k_x\,{\bf e}_x$,
directed  along the $x$ axis. In Fig.~1a) and b) the transverse
magnetic (TM) and the transverse electric (TE) modes in the cavity
corresponding to this $\bf k$ vector are indicated \cite{fotnote}.

Our conventions above mean that we can let $\nabla^2 \rightarrow
\partial_z^2-k_x^2$. We can now write down the governing equations
for the Fourier components $g_{ik}$, from Eqs.~(\ref{14}). The
simplest equation follows by setting $(ik)=(yy)$:
\begin{equation}
 \partial_z^2 g_{yy}-( \kappa^2-\frac{2\mu k_x \omega}{c}\,\chi_{yz}
)g_{yy}=-\frac{\mu\omega^2}{c^2}\delta(z-z'), \label{22}
\end{equation}
where we have defined
\begin{equation}
\kappa^2=k_x^2-\varepsilon \mu \omega^2/c^2. \label{23}
\end{equation}
Equation (\ref{22}) is uncoupled; this being a consequence of our
choice for $\bf k$ implying that $\partial_y \rightarrow 0$.

Setting $(ik)=(xx)$ we obtain
\begin{equation}
i(k_x+\frac{\mu \omega}{c}\chi_{zy})\partial_z
g_{zx}-(\partial_z^2+\frac{\varepsilon \mu
\omega^2}{c^2})g_{xx}=\frac{\mu \omega^2}{c^2}\delta(z-z'),
\label{24}
\end{equation}
and with $(ik)=(zz)$,
\begin{equation}
i(k_x+\frac{\mu\omega}{c}\chi_{zy})\partial_zg_{xz}+(\kappa^2+\frac{2\mu
k_x\omega}{c}\,\chi_{zy})g_{zz}=\frac{\mu
\omega^2}{c^2}\,\delta(z-z'). \label{25}
\end{equation}
The last two equations are coupled. Consider finally the
nondiagonal components: with $(ik)=(zx)$ we obtain
\begin{equation}
i(k_x+\frac{\mu\omega}{c}\chi_{zy})\partial_z
g_{xx}+(\kappa^2+\frac{2\mu k_x\omega}{c}\chi_{zy})g_{zx}=0,
\label{26}
\end{equation}
and with $(ik)=(xz)$,
\begin{equation}
i(k_x+\frac{\mu\omega}{c}\chi_{zy})\partial_z g_{zz}-(\partial_z^2
+\frac{\varepsilon\mu\omega^2}{c^2})g_{xz}=0. \label{27}
\end{equation}
The coupling in the differential equation (\ref{24}) for $g_{xx}$
can be removed if we make use of Eq.~(\ref{26}) differentiated
with respect to $x$. Some manipulations, again observing that
$\chi_{ik}$ is small, yield
\begin{equation}
\partial_z^2\,g_{xx}-K^2 g_{xx}=\frac{K^2}{\varepsilon}\delta(z-z'),
\quad K=\kappa (1+\frac{\mu k_x \omega}{\kappa^2
c}\chi_{zy}).\label{28}
\end{equation}
Equation (\ref{22}) can be rewritten similarly:
\begin{equation}
\partial_z^2\, g_{yy}-L^2 g_{yy}=-\frac{\mu \omega^2}{c^2}\delta
(z-z'), \quad L= \kappa (1-\frac{\mu k_x \omega}{\kappa^2
c}\chi_{yz}). \label{29}
\end{equation}
The differential equations  (\ref{28}) and (\ref{29}) for the
diagonal components are convenient for further manipulation. Note
that the values of $K$ and $L$ are dependent on whether the
direction of propagation of the wave is to the right or to the
left. If $\chi_{ik}=0$, the expressions agree with those of
Ref.~\cite{ellingsen07} \cite{fotnote2}.

We now proceed to solve the equations, beginning with
Eq.~(\ref{29}). As $E_y=0$ at $z=0$ and $z=a$ because of the
boundary conditions, we  have $g_{yy}(0,z',{\bf k},
\omega)=g_{yy}(a, z', {\bf k}, \omega)=0$. The solution of
Eq.~(\ref{29}) can then be written
\begin{equation}
g_{yy}=\frac{\mu\omega^2}{2Lc^2}\left\{
e^{-L|z-z'|}-e^{-L(z+z')}+\frac{ 2[\cosh L(z-z')-\cosh
L(z+z')]}{\exp{(2La)}-1}\right\}. \label{30}
\end{equation}

When $\chi_{ik}=0$, this expression agrees with that given in
Appendix C of Ref.~\cite{hoye03} in the limit of perfectly
conducting plates.

Next considering $g_{xx}$, we must analogously have $g_{xx}(0, z',
{\bf k}, \omega)=g_{xx}(a, z', {\bf k}, \omega)=0$ in view of the
boundary conditions. The solution of Eq.~(\ref{28}) becomes
\begin{equation}
g_{xx}=-\frac{K}{2\varepsilon}\left\{
e^{-K|z-z'|}-e^{-K(z+z')}+\frac{2[\cosh K(z-z')-\cosh
K(z+z')]}{\exp{(2Ka)}-1}\right\}, \label{31}
\end{equation}

The expressions (\ref{30}) and (\ref{31}) are fairly complicated.
For practical purposes it is  possible to simplify the expressions
considerably, by omitting terms containing $(z+z')$. The reason is
that these terms do not contribute to physical quantities like the
Casimir force on the plates or to the field momentum in the gap.
This can be seen in two different ways. The simplest way is to
argue, as in Sect. 81 in \cite{lifshitz80}, that by putting $z=z'$
in solutions having the argument $(z+z')$ one would obtain
physical quantities like field momentum in the gap varying with
the position $z$. This would contradict the law of conservation of
momentum. Another way of examining this rather subtle point is to
include the $(z+z')$ terms everywhere in the formalism, and to
verify that they really do not contribute in the end. In addition
to the discussion in  \cite{lifshitz80}, one can find more
mathematical details about this point in the paper
\cite{ellingsen07} and in the thesis \cite{ellingsen06}.

We can moreover omit the source-dependent inhomogeneous $|z-z'|$
term in each of  the Green functions. This term represents the
solution  pertaining to the delta function source inside a
homogeneous medium filling all space.  Being geometry independent,
it cannot contribute to any physical quantity related to the
geometry. All in all, we shall in the following use  the
"effective" Green functions
\begin{equation}
g_{xx}=-\frac{K}{\varepsilon}\,\frac{\cosh
K(z-z')}{\exp{(2Ka)}-1}, \label{32}
\end{equation}
\begin{equation}
g_{yy}=\frac{\mu\omega^2}{Lc^2}\,\frac{\cosh
L(z-z')}{\exp{(2La)}-1}. \label{33}
\end{equation}
Consider finally the remaining diagonal component, $g_{zz}$. To
this end we first observe the symmetry property
\begin{equation}
g_{xz}(z,z',{\bf k}, \omega)=g_{zx}(z',z,-{\bf k},\omega),
\label{34}
\end{equation}
which is an example of the general relation
\begin{equation}
\Gamma_{ik}({\bf r, r'},\tau)=\Gamma_{ki}({\bf r', r},-\tau),
\label{35}
\end{equation}
when expressed in  Fourier space (cf. Sect. 81 in
Ref.~\cite{lifshitz80}).
 From Eq.~(\ref{26}) we have, when inserting the expression
 (\ref{32}),
 \begin{equation}
 g_{zx}(z,z',{\bf k},
 \omega)=\frac{i}{\varepsilon}(k_x+\frac{\mu\omega}{c}\chi_{zy})\,\frac{\sinh
 K(z-z')}{\exp {(2Ka)}-1}.  \label{36}
 \end{equation}
 Now, to the required order $K(-k_x)=\kappa^2/K(k_x)$, according
 to Eq.~(\ref{28}). Thus we get from Eq.~(\ref{34})
 \begin{equation}
 g_{xz}(z,z',{\bf k},
 \omega)=\frac{i}{\varepsilon}(k_x-\frac{\mu\omega}{c}\chi_{zy})\,\frac{\sinh[\kappa^2
 (z-z')/K]}{\exp (2\kappa^2 a/K)-1}, \label{37}
 \end{equation}
where here and henceforth $K=K(k_x)$ as defined in Eq.~(\ref{28}).
From Eq.~(\ref{25}) we then finally  get (delta-function omitted):
\begin{equation}
g_{zz}=\frac{\kappa^2 k_x^2}{K^3
\varepsilon}\,\frac{\cosh[\kappa^2 (z-z')/K]}{\exp(2\kappa^2
a/K)-1}. \label{38}
\end{equation}
These Green-function expressions have to our knowledge not been
derived before. Before applying them to the Feigel effect, we
shall in the next section follow a more simplistic approach and
consider the right/left asymmetry in the field momentum
considering one single mode only.

\section{Energy-momentum formalism. Right/left field momentum
asymmetry}

\subsection{Energy-momentum formalism}

Before considering the momentum asymmetry for one single chosen
direction of propagation, we need to develop the formalism related
to the electromagnetic energy-momentum tensor. In this subsection
we take a general approach, allowing for external charges $\rho$
and currents $\bf J$. The coupling tensor $\chi_{ik}$ is allowed
to be general (not necessarily of the form given in
Eq.~(\ref{21})), though constant, and we allow also for optical
anisotropy by letting $\varepsilon \,\delta_{ik} \rightarrow
\varepsilon_{ik}$, $\mu \,\delta_{ik}\rightarrow \mu_{ik}$ with
the material parameters constant.

It is convenient to write the constitutive relations (\ref{1}) and
(\ref{2}) on tensor form,
\begin{equation}
D_i=\varepsilon_0\varepsilon_{ik}E_k+\frac{1}{\mu_0c}\chi_{ik}B_k,
\label{39}
\end{equation}
\begin{equation}
H_i=-\frac{1}{\mu_0c}\chi_{ki}\,E_k+\frac{1}{\mu_0}{\mu^{-1}}_{ik}\,
B_k. \label{40}
\end{equation}
In view of Maxwell's equations (\ref{5}) and (\ref{6}) we obtain
the conservation equation for energy,
\begin{equation}
{\bf \nabla \cdot S}+\dot w=-{\bf E\cdot J}, \label{41}
\end{equation}
where ${\bf E\cdot J}$ is the energy dissipation,
\begin{equation}
\bf S=E\times H \label{42}
\end{equation}
the Poynting vector, and
\begin{equation}
w=\frac{1}{2}(\bf E\cdot D+H\cdot B) \label{43}
\end{equation}
the energy density.

As for the momentum conservation, it is convenient to start from
the equation
\begin{equation}
\partial_t({\bf D\times B})_i=-\rho
E_i-\epsilon_{ijk}J_j B_k+\partial_k(E_i D_k+H_i
B_k)-\varepsilon_{kl}E_{k,i}E_l-{\mu^{-1}}_{kl}B_{l,i}B_k,
\label{44}
\end{equation}
which follows from Maxwell's equations (here $E_{i,k}\equiv
\partial_k E_i$, etc.). Introducing the Lorentz force density
\begin{equation}
{\bf f}^L=\rho {\bf E+J\times B}, \label{45}
\end{equation}
as well as the Minkowski stress tensor \cite{minkowski10},
 \begin{equation}
 T_{ik}^M=E_iD_k+H_iB_k-\frac{1}{2}(\bf E\cdot D+H\cdot B),
 \label{46}
 \end{equation}
 we can write the momentum conservation equation as
 \begin{equation}
 \partial_kT_{ik}^M-\dot{g}_i^M=f^L_i, \label{47}
 \end{equation}
 where
 \begin{equation}
 {\bf g}^M=\bf D\times B \label{48}
 \end{equation}
 is the Minkowski momentum density. (The symbol $\bf g$ for momentum is not to be confused
 with the Green functions.) It is generally known that the
 above expressions hold when the medium is optically anisotropic.
 It is however somewhat remarkable that they hold when $\chi_{ik}
 \neq 0$ also; there seems to be no simple physical reason why $\chi_{ik}$
 should  drop out from the formalism.

 In the case of high frequency fields, in particular optical
 fields, the Minkowski theory appears to be both simple and
 capable of describing all experiments (cf. the analysis of one of
 the present authors on this point some years ago \cite{brevik79};
  some more recent papers
  are listed in Ref.~\cite{ellingsen07}). However, at low
  frequencies where the effect of the oscillations are themselves
  observable - notably in the
  Lahoz-Walker experiment \cite{walker75} - the experiments agree
  not with the Minkowski but rather with the Abraham force, which
  accordingly can be taken to be the most 'physical' alternative
  at these frequencies. The Abraham theory \cite{abraham10} consists in
  symmetrizing the stress tensor,
  \begin{equation}
  T_{ik}^A=\frac{1}{2}(E_iD_k+E_kD_i)+\frac{1}{2}(H_iB_k+H_kB_i)-\frac{1}{2}({\bf
  E\cdot D+H\cdot B}),\label{49}
  \end{equation}
  and taking the momentum density to be
   \begin{equation}
   {\bf g}^A=\frac{1}{c^2}\bf E\times H, \label{50}
   \end{equation}
   the latter satisfying the relation ${\bf g=S}/c^2$, the
   so-called Planck's principle of inertia of energy.

   We assume henceforth optical anisotropy so that $\varepsilon$
   and $\mu$ are scalars, and also that $\rho=0, \,{\bf J}=0$. The
   Minkowski and Abraham stress tensors become thereby equal,
   $T^M_{ik}=T^A_{ik}$. The Abraham conservation equation for
   momentum can be written as
   \begin{equation}
   \partial_k T^A_{ik}-\dot{g}^A_i=[(\varepsilon
   \mu-1)/c^2]\partial_t({\bf E\times H})_i, \label{51}
   \end{equation}
   where the term on the right hand side is the 'Abraham term'. It
   was precisely this term that was measured by Walker and Lahoz
   \cite{walker75}. In a high-frequency field, it fluctuates out.

   We shall return to the Abraham force in  Sect. 6.

   \subsection{Momentum asymmetry}

   Referring to Fig.~1, we consider to begin with only the
   right-moving TE wave corresponding to the field components
   \begin{equation}
   E_y=\sqrt{\frac{2}{a}}\,\sin k_n z\,
    e^{i({\bf k\cdot x}-\omega
   t)},
   \label{52}
   \end{equation}
   \begin{equation}
   B_x=\sqrt{\frac{2}{a}}\,\frac{ik_n}{\omega}\cos k_n z\,e^{i({\bf k\cdot x}-\omega
   t)}, \label{53}
   \end{equation}
   \begin{equation}
   B_z=\sqrt{\frac{2}{a}}\,\frac{k_x}{\omega}\sin k_n z\,e^{i({\bf k\cdot x}-\omega
   t)}\label{54}
   \end{equation}
   the other components being zero. (We use the same normalization of the fields as Tiggelen
   {\it et al.}\cite{tiggelen06}.) Here ${\bf k\cdot x}=k_x x$,
   and $ k_n=\pi n/a$ with $n=1,2,3...$ is the transverse wave
   number. For a given value of $k_x$, the eigenfrequencies
   $\omega$ are thus discrete. We can derive the dispersion
   equation by going back to the field equation (\ref{9}) for
   $E_i=E_y$ in the source-free case, observing that
   $\partial_lE_l=\partial_yE_y=0$, inserting the form (\ref{21})
   for $\chi_{ik}$. We obtain
   \begin{equation}
   \left(k_x^2+k_n^2-\frac{\varepsilon \mu}{c^2}\omega^2-\frac{2\mu
   k_x\omega}{c}\chi_{yz}\right)E_y=0, \label{55}
   \end{equation}
   which implies to the lowest order in $\chi_{yz}$
   \begin{equation}
   \omega=\frac{c}{\sqrt{\varepsilon \mu}}\sqrt{k_x^2+k_n^2}\left[
   1-\sqrt{\frac{\mu }{\varepsilon}}
   \,\frac{k_x}{\sqrt{k_x^2+k_n^2}}\,\chi_{yz} \right]. \label{56}
   \end{equation}
The right/left asymmetry is manifest. A left-moving wave is
described by the substitution $k_x\rightarrow -k_x$.

Let us now calculate the field energy density, $w$, for the TE
mode. We get
\[ w=\frac{1}{4}({\bf E\cdot D^*+H\cdot B^*)} \]
\begin{equation}
=\frac{\varepsilon_0\varepsilon}{2a}\left[1+\frac{c^2}{\varepsilon
\mu}\,\frac{k_x^2}{\omega^2}\right] \sin^2 k_n
z+\frac{1}{2\mu_0\mu a}\,\frac{k_n^2}{\omega^2}\cos^2k_nz;
\label{57}
\end{equation}
the $\chi_{yz}$ terms drop out when $w$ is written in this way. It
is convenient to consider the expression integrated from $z=0$ to
$z=a$, thereby getting the energy $W$ per unit length and width,
\begin{equation}
W=\int_0^a w dz=\frac{\varepsilon_0\varepsilon}{4}\left[
1+\frac{c^2}{\varepsilon \mu}\,\frac{k_x^2+k_n^2}{\omega^2}
\right]. \label{58}
\end{equation}
Using Eq.~(\ref{56}) we can write this in terms of the wave number
components,
\begin{equation}
W=\frac{\varepsilon_0\varepsilon}{2}\left[
1+\frac{1}{2}\sqrt{\frac{\mu}{\varepsilon}}\,\frac{k_x}{\sqrt{k_x^2+k_n^2}}\,\chi_{yz}\right].
\label{59}
\end{equation}
The Poynting vector in the $x$ direction, $S_x$, may be calculated
as
\begin{equation}
S_x=\frac{1}{2}({\bf E\times H^*})_x=\left( \frac{1}{\mu_0\mu
a}\,\frac{k_x}{\omega}-\frac{\chi_{yz}}{\mu_0ca}\right) \sin^2 k_n
z, \label{60}
\end{equation}
which means that the integrated energy flux when expressed in
terms of wave number becomes
\begin{equation}
q_x=\int_0^a S_x dz=\frac{\varepsilon_0 c}{2}
 \left[
\sqrt{\frac{\varepsilon}{\mu}}\,\frac{k_x}{\sqrt{k_x^2+k_n^2}}-\frac{k_n^2}{k_x^2+k_n^2}\,\chi_{yz}
\right]. \label{61}
\end{equation}
Alternatively, we might calculate the energy flux as $q_x=Wu_x$,
where $u_x$ is the group velocity
\begin{equation}
u_x=\frac{\partial \omega}{\partial
k_x}=\frac{c}{\sqrt{\varepsilon
\mu}}\,\frac{k_x}{\sqrt{k_x^2+k_n^2}}-\frac{c}{\varepsilon}\chi_{yz}.
\label{62}
\end{equation}
This agreement is as we should expect, since we are dealing with
the propagation of low-amplitude waves. The kinematic group
velocity concept and the dynamic energy flow velocity concept
should be the same.

Consider finally the Minkowski momentum density $g_x^M$:
\begin{equation}
g_x^M=\frac{1}{2}({\bf D\times
B^*})_x=\frac{\varepsilon_0\varepsilon}{a}\frac{k_x}{\omega}\left(
1+\frac{k_xc}{\varepsilon \omega}\,\chi_{yz}\right)\sin^2 k_nz.
\label{63}
\end {equation}
Comparison between Eqs.~(\ref{60}) and (\ref{63}) shows that the
relationship $g_x^M=(\varepsilon\mu/c^2)S_x$, known from
conventional optics, does {\it not} hold when $\chi_{yz}$ is
different from zero. We also give the expression (\ref{63}) when
integrated over $z$:
\begin{equation}
G_x^M=\int_0^ag_x^M
dz=\frac{\varepsilon_0\varepsilon}{2}\frac{k_x}{\omega}\left(1+\frac{k_xc}{\varepsilon
\omega}\,\chi_{yz}\right). \label{64}
\end{equation}
Again, the right/left asymmetry is manifest.

\section{The Green-function approach to the Feigel effect}

Our intention now is to calculate the Minkowski momentum asymmetry
in the chiral medium using the Green-function approach from
Section 3. We start from the following general expression,
reverting to real representation for the fields,
\begin{equation}
{\bf g}^M=\lim_{ x' \rightarrow x}\int_{-\infty}^\infty
\frac{d\omega}{2\pi}e^{-i\omega \tau}\int \frac{d^2
k}{(2\pi)^2}\,e^{i\bf k\cdot(r-r')}\,\langle {\bf D(r)\times
B(r')}\rangle_{\omega \bf k} \label{65}
\end{equation}
We  assume zero temperature, so that the brackets
$\langle\,\,\rangle$ mean purely quantum mechanical average. As no
thermal fields are excited, the field momentum as well as the
field energy stem exclusively from the vacuum zero-point
oscillations. Whereas in the previous section we considered the
contribution from one single selected mode only, we shall now
consider the effect of summing over all available vacuum modes. We
shall impose one restriction, however: the wave number $\bf k$
will be required to lie either in the positive or the negative $x$
direction. This corresponds to our Green-function approach in
Sect.~3. Mathematically, it means that we can let $\int d^2
k/(2\pi)^2 \rightarrow \int dk_x/2\pi$. As the distribution of
fields  does not vary in the transverse $y$ direction, we can
effectively let $\partial_y \Rightarrow 0$  when applied to the
fields. Evidently, the $x$ component of field momentum has to be
zero in the case of a non-chiral medium; if there is an asymmetry
following from the formalism this has to be caused by the presence
of $\chi_{ik}$. As before, we assume the particular form
(\ref{21}) for $\chi_{ik}$.

The $x$ component of Eq.~(\ref{65}) becomes (we omit the 'lim'
from now on)
\begin{equation}
g_x^M=\int_{-\infty}^\infty \frac{d\omega}{2\pi} e^{-i\omega
\tau}\,\int_{-\infty}^\infty
\frac{dk_x}{2\pi}e^{ik_x(x-x')}\langle D_y({\bf r})B_z({\bf
r'})-D_z({\bf r})B_y({\bf r'})\rangle_{\omega k}. \label{66}
\end{equation}
We insert from Eqs.~(\ref{1}) and (\ref{2})
\begin{equation}
D_y=\varepsilon_0\varepsilon E_y+\frac{\chi_{yz}}{\mu_0 c}B_z,
\label{67}
\end{equation}
\begin{equation}
D_z=\varepsilon_0\varepsilon E_z+\frac{\chi_{zy}}{\mu_0 c}B_y,
\label{68}
\end{equation}
and get
\[ g_x^M=  \int_{-\infty}^\infty \frac{d\omega}{2\pi} e^{-i\omega
\tau}\,\int_{-\infty}^\infty \frac{dk_x}{2\pi}e^{ik_x(x-x')}
 \Big[ \varepsilon_0 \varepsilon \langle E_y({\bf r})B_z({\bf
 r'})\rangle_{\omega k} \]
\begin{equation}
-\varepsilon_0\varepsilon \langle E_z({\bf r})B_y({\bf
r'})\rangle_{\omega k}-\frac{\chi_{zy}}{\mu_0 c}\langle B_y({\bf
r})B_y({\bf r'})\rangle_{\omega k}+\frac{\chi_{yz}}{\mu_0c}\langle
B_z({\bf r})B_z({\bf r'})\rangle_{\omega k} \Big]. \label{69}
\end{equation}
We have thus so far expressed $g_x^M$ in terms of the two-point
functions for the fundamental fields. Using Eqs.~(\ref{16}) and
(\ref{17}) we calculate
\begin{equation}
\langle E_y({\bf r})B_z({\bf r'})\rangle_{\omega
k}=\frac{\hbar}{\varepsilon_0}\,\frac{-ik_x}{\omega} \,{\rm Im}\,
g_{yy}, \label{70}
\end{equation}
\begin{equation}
\langle E_z({\bf r})B_y({\bf r'})\rangle_{\omega
k}=\frac{\hbar}{\varepsilon_0}\,\frac{1}{\omega}(\partial_z'\,{\rm
Im}\,g_{zx}+ik_x\,{\rm Im}\,g_{zz}), \label{71}
\end{equation}
\[ \langle B_y({\bf r})B_y({\bf r'})\rangle_{\omega
k}=\frac{\hbar}{\varepsilon_0}\,
\frac{i}{\omega^2}(\partial_z^2\,{\rm Im}\,g_{xx}-ik_x\,{\rm
Im}\,\partial_z g_{zx} \]
\begin{equation}
-ik_x\partial_z\,{\rm Im}\,g_{xz}-k_x^2\, {\rm Im}\,  g_{zz}),
\label{72}
\end{equation}
\begin{equation}
\langle B_z({\bf r})B_z({\bf r'})\rangle_{\omega
k}=\frac{\hbar}{\varepsilon_0}\,\frac{-ik_x^2}{\omega^2}\,{\rm
Im}\, g_{yy}. \label{73}
\end{equation}
We have here, as above, naturally defined
$\langle\,\,\rangle_{\omega k}$ via the relation
\begin{equation}
 \langle E_y({\bf r})B_z({\bf r'})\rangle_{\omega
}= \int_{-\infty}^\infty \frac{dk_x}{2\pi}\,e^{ik_x(x-x')}\,
\langle E_y({\bf r})B_z({\bf r'})\rangle_{\omega k}, \label{74}
\end{equation}
etc. We can thus express $g_x^M$ as
\begin{equation}
g_x^M=\hbar \int_{-\infty}^\infty \frac{d\omega}{2\pi
\omega}\,e^{-i\omega \tau}\,\int_{-\infty}^\infty
\frac{dk_x}{2\pi}\,e^{ik_x(x-x')}\,\langle\,\,\rangle, \label{75}
\end{equation}
where
\[ \langle\,\,\rangle =-ik_x\varepsilon \,{\rm
Im}\,g_{yy}-\varepsilon (\partial_z'\,{\rm Im}\,g_{zx}+ik_x\,{\rm
Im}\, g_{zz}) \]
\[
+\frac{ic}{\omega}\chi_{zy}(-\partial_z^2\,{\rm Im}\,
\,g_{xx}+ik_x\, {\rm Im}\,\partial_zg_{zx}\]
\begin{equation}
+ik_x\partial_z \,{\rm Im}\,g_{xz}+k_x^2\,{\rm Im}\, g_{zz})
-\frac{ik_x^2c}{\omega}\chi_{yz}\,{\rm Im}\, g_{yy}. \label{76}
\end{equation}
This expression shows that it is necessary to calculate $g_{yy},
g_{zx}$ and $g_{zz}$ to order $\chi_{ik}$. From Eqs.~(\ref{33}),
(\ref{36}) and (\ref{38}) we get
\begin{equation}
g_{yy}=\frac{\mu \omega^2}{c^2}\frac{1}{\kappa d}\left[
1+\frac{\mu k_x\omega}{\kappa^2c}\left(1+\frac{2\kappa
a}{d}e^{2\kappa a}\right)\chi_{yz}\right], \label{77}
\end{equation}
\begin{equation}
 \partial_z' g_{zx}=-\partial_z g_{zx}
=-\frac{i\kappa k_x}{\varepsilon d}\left\{ 1+\frac{\mu
\omega}{k_xc}\left[1+\frac{k_x^2}{\kappa^2}\left(1-\frac{2\kappa
a}{d} e^{2\kappa a}\right)\right]\chi_{zy}\right\}, \label{78}
\end{equation}
\begin{equation}
g_{zz}=\frac{k_x^2}{\varepsilon \kappa d}\left[ 1-\frac{3\mu
k_x\omega}{\kappa^2c}\left(1- \frac{2\kappa a}{3}\frac{e^{2\kappa
a}}{d}\right)\chi_{zy}\right], \label{79}
\end{equation}
where $d$ is defined as
\begin{equation}
d=e^{2\kappa a}-1. \label{80}
\end{equation}
The remaining terms in Eq.~(\ref{76}) are however multiplying
$\chi_{zy}$ or $\chi_{yz}$, and so need not to be expanded in
$\chi_{ik}$. Thus to sufficient accuracy
\begin{equation}
\partial_z^2 g_{xx}=-\frac{\kappa^3}{\varepsilon}\,\frac{1}{d},
\label{81}
\end{equation}
\begin{equation}
\partial_zg_{zx}=\partial_zg_{xz}=\frac{i\kappa
k_x}{\varepsilon}\,\frac{1}{d}, \label{82}
\end{equation}
\begin{equation}
g_{yy}=\frac{\mu \omega^2}{\kappa c^2}\,\frac{1}{d}, \label{83}
\end{equation}
\begin{equation}
g_{zz}=\frac{k_x^2}{\kappa \varepsilon}\,\frac{1}{d}. \label{84}
\end{equation}
We now put $\tau=0,\, x-x'=0$ in Eq.~(\ref{75}), and perform a
standard complex frequency rotation whereby $\omega \rightarrow
i\zeta$, with $\zeta$ real \cite{schwinger78}. As $d\omega/\omega
\rightarrow d\zeta/\zeta$, it follows from Eq.~(\ref{75}) that of
physical importance are only those terms in $\langle \,\,\rangle$
that are real after the rotation ($g_x^M$ has to be real). Thus
the first terms in Eqs.~(\ref{77}), (\ref{78}) and (\ref{79}) do
not contribute. This is  what we should expect: the asymmetry in
momentum is caused by $\chi_{ik}$. After some calculation we
obtain, by  letting $\int_{-\infty}^\infty d\zeta \rightarrow
2\int_0^\infty d\zeta$, $\int_{-\infty}^\infty dk_x \rightarrow
2\int_0^\infty dk_x$ because of symmetry of the integrand about
the origin,
\[ g_x^M=\frac{4\hbar \mu}{c}\int_0^\infty
\frac{d\zeta}{2\pi}\int_0^\infty
\frac{dk_x}{2\pi}\frac{k_x^4}{\kappa^3
d}\Bigg\{\left[1-\frac{\varepsilon
\mu\zeta^2}{k_x^2c^2}\frac{2\kappa a }{d}e^{2\kappa
a}\right]\chi_{yz} \]
\begin{equation}
-\frac{2\kappa^2}{k_x^2}\left[
1+\frac{3k_x^2}{2\kappa^2}-\frac{\kappa
a}{d}\left(1+\frac{k_x^2}{\kappa^2}\right)e^{2\kappa
a}\right]\chi_{zy}\Bigg\}. \label{85}
\end{equation}
Recall that $d$ is given by Eq.~(\ref{80}), where now
$\kappa^2=k_x^2+\varepsilon \mu \zeta^2/c^2$. The integrals are
seen to be finite. This is so because we have already performed
the regularization by omitting those parts in the Green function
that refer to the infinite undisturbed system. (Cf. also the
remarks at the end of Sect.~1.) If the separation becomes
infinite, then $d\rightarrow e^{2\kappa a} \rightarrow \infty$,
and $g_x^M \rightarrow 0$ as it must; all plate-induced physical
effects have to go away in this limit.

The expression (\ref{85}) may be conveniently rewritten in terms
of polar coordinates. Introduce $X=k_x=\kappa \cos
\theta,\,Y=(\sqrt{\varepsilon \mu}/c) \zeta =\kappa \sin \theta$,
so that
\begin{equation}
X^2+Y^2=\kappa^2. \label{86}
\end{equation}
The area element in the $XY$ plane is $\kappa d\kappa
d\theta=(\sqrt{\varepsilon \mu}/c)dk_xd\zeta$. Then
\[
g_x^M=\frac{\hbar}{\pi^2}\sqrt{\frac{\mu}{\varepsilon}}\int_0^{\pi/2}\cos^4
\theta d\theta\int_0^\infty \frac{\kappa^2
d\kappa}{d}\Bigg\{\left[1-\tan^2\theta \,\frac{2\kappa
a}{d}e^{2\kappa a}\right]\chi_{yz} \]
\begin{equation}
-\left[5+2\tan^2\theta -(2 +\tan^2\theta)\frac{2\kappa
a}{d}e^{2\kappa a}\right]\chi_{zy}\Bigg\}. \label{87}
\end{equation}
The integrals can be evaluated to give
\begin{equation}
g_x^M=\frac{\hbar \zeta(3)}{16\pi
a^3}\sqrt{\frac{\mu}{\varepsilon}}\,\chi_{zy}, \label{88}
\end{equation}
where $\zeta(3)$ is the Riemann zeta function with argument 3. It
is noteworthy that only one of the ME coefficients, $\chi_{zy}$,
appears in this expression. The factor multiplying $\chi_{yz}$ in
Eq.~(\ref{87})  turns out to be zero. There seems to be no simple
reason for this, although the behavior is obviously related to the
complicated structure of Eq.~(\ref{69}) and the need to expand
$g_{yy}, g_{zx}$ and $g_{zz}$ in $\chi_{ik}$; cf.
Eqs.~(\ref{77})-(\ref{79}). Recall that the $x$ direction has been
singled out as special, and also that we have taken the variation
of the fields in the transverse $y$ direction to be equal to zero.
To our knowledge, an expression of this kind has not been derived
before.

The quantity $g_x^M$ is measurable, in principle. Before any
measurement can be done, the expression (\ref{88}) has of course
to be augmented by  contributions from all the other values of
$\bf k$. Aspects connected with real experiments lie outside the
scope of the present paper.

\section{Summary, and discussion}

Let us first recall the assumption on which the above calculation
is based:

1.   The tensor $\chi_{ik}$ characterizing the magnetoelectric
medium is given naturally over all space in the fluid, on the
inside as well as on the outside of the conducting plates. The
constitutive relations are Eqs.~(\ref{1}) and (\ref{2}), or, in
inverse form, Eqs.~(\ref{3}) and (\ref{4}) when the magnitude
$|\chi_{ik}|$ of the coupling is small. In the example that we
calculated in detail, $\chi_{ik}$ is given by Eq.~(\ref{21}). The
tensor $\chi_{ik}$ may be asymmetric, in contrast to the
permittivity $\varepsilon_{ik}$ and permeability $\mu_{ik}$ which
are always symmetric.

2.  We have followed two different approaches, giving most weight
to the Green-function approach since this does not seem to be
treated very much in the literature.  We took the temperature to
be zero. The governing equation for the dyad $\Gamma_{ik}$ is
Eq.~(\ref{14}). The full solution of two of the diagonal
components, $g_{yy}$ and $g_{xx}$, introduced as Fourier
components of the $\Gamma$'s via Eq.~(\ref{19}), are given by
Eqs.~(\ref{30}) and (\ref{31}). We have here assumed that there is
no variation of the fields in the transverse $y$ direction. The
derivation of Eqs.~(\ref{30}) and (\ref{31}) generalizes
conventional Green-function Casimir theory \cite{milton01,hoye03}
to the case of ME media. For practical purposes it turns out to be
possible to simplifying the expressions considerably, by omitting
terms that do not contribute to physical quantities in the end.
The arguments for proceeding in this way are spelled out, for
instance,  in Ref.~\cite{lifshitz80}. The relevant reduced
components of $g_{ik}$ in our case are given at the end of
Sect.~3.

3.  In Sect.~4 we deviated to follow a different, and more simple,
approach. After having established the momentum conservation
equation  for a ME medium, we calculated the right/left asymmetry
for one single mode only (choosing one of the modes considered in
Ref.~\cite{tiggelen06}). The results are given by Eqs.~(\ref{63})
and (\ref{64}). Adding two similar modes, one propagating in the
$+x$ direction and one in the $-x$ direction, we obtain a net flow
of momentum, caused by the coupling $\chi_{yz}$.

4.  In Sect.~5 we returned to the Green-function approach,
calculating the net $x$ component of momentum arising now not from
one single mode, but from all modes propagating in the $\pm x$
directions in the {\it vacuum} field. The main result is given by
Eq.~(\ref{88}). All terms independent of $\chi_{ik}$ drop
automatically out of the formalism, in accordance with what we
should expect beforehand.

5.  On physical grounds one may ask: where does the net
electromagnetic momentum come from? Obviously, it cannot come from
'nothing'.  We are actually comparing two different physical
situations here. The first is when the conducting plates are
infinitely far separated. This is our initial 'vacuum' state. The
final state is when the plates have been brought close to each
other, infinitely slowly. The calculated quantity $g_x^M$ is the
Minkowski momentum density extracted during this process of change
of the plate separation. The coupling parameter $\chi_{ik}$ in the
fluid is the same, all the time. The process is thus conceptually
quite close to the process encountered in usual Casimir theory;
the main difference being that it is now momentum, not energy,
that is extracted.

6.  The setting of our thought experiment is similar, but not
exactly the same, as that envisaged in Feigel's paper
\cite{feigel04}. Feigel assumed the coupling $\chi_{ik}$ in the
fluid to be  the result of applying strong electric and magnetic
fields. We have deliberately avoided this picture  since it
complicates the situation in the sense that  one  has to deal with
two sets of fields, both the external fields, and the wave modes.
When assuming naturally occurring $\chi_{ik}$ instead, as we have
done, the interpretation of the effect becomes more transparent.

Before leaving this idea, let us however not the following point:
Assume that strong crossed fields ${\bf E}_0$ and ${\bf H}_0$ are
applied between the conducting plates at the instant $t=0$. Then,
during the  time when the external fields increase in strength,
there acts an Abraham force in the fluid in the interior. The
force density is given by the expression on the right in
Eq.~(\ref{51}). Integrating over time, from $t=0$ until the
external fields have become constant, we see that the following
mechanical momentum density is imparted to the fluid:
\begin{equation}
{\bf g}^A=\frac{\varepsilon \mu -1}{c^2}({\bf E}_0\times {\bf
H}_0). \label{89}
\end{equation}
This is the dominant momentum given to the fluid between the
plates. In addition comes the momentum transferred from the wave
modes; these are connected with $\chi_{ik}$. The momentum
(\ref{88}) is actually very similar to the momentum (or more
strictly the angular momentum) transferred to the suspended
dielectric cylindrical shell in the  Walker-Lahoz experiment in
ordinary electrodynamics \cite{walker75,brevik79}.

7. It might appear surprising that in Feigel's paper a
high-frequency cutoff $\omega_{cut}$ is introduced, whereas in the
present treatment there is no need of a cutoff. The reason for
this behavior is that the two formalisms are constructed
differently: Feigel considers the total contribution, including
that of the infinite unconstrained system, whereas in our case we
have regularized the infinite  contribution away. The case of high
frequencies leads in Feigel's case to infinities, whereas in our
case it leads to {\it zero}. Again, this is the same point as was
emphasized at the end of Sect.~1. We are generally looking at the
present problem as a sort of  Casimir-type problem.

8.  What is the connection  between the Feigel effect and
relativity? In this context it might be of interest to recall how
the relativistic formulation of electrodynamics in continuous
media is formulated. There is always one particular inertial
system  $S^0$ here, namely the one where the medium is at rest -
this was emphasized already in the classic papers of Jauch and
Watson \cite{jauch48}. The relativistic formulation is obtained by
introducing two electromagnetic field tensors $F_{\mu\nu}$ and
$H_{\mu\nu}$ such that the covariant Maxwell equations
\begin{equation}
 \partial_\rho F_{\mu\nu}+\partial_\mu F_{\nu\rho}+\partial_\nu
F_{\rho\mu}=0, \quad \partial_\nu H_{\mu\nu}=0 \label{90}
\end{equation}
agree with the standard Maxwell equations in $S^0$ (we assume no
external charges or currents). The electromagnetic energy-momentum
tensor $S_{\mu\nu}$, assuming Minkowski's expression for the
momentum density, is divergence-free,
\begin{equation}
\partial_\nu S_{\mu\nu}=0, \label{91}
\end{equation}
meaning that the energy and momentum of the total field constitute
a four-vector. Moreover, this four-vector is {\it space-like}, so
that it is possible to find inertial systems where the radiation
energy becomes negative. A striking demonstration of this property
is found in connection with the Cherenkov effect, in the frame
where the emitting particle is at rest. A clear introduction to
this kind of theory is found in M{\o}ller's book \cite{moller72},
and the theory is discussed also  in papers of one of the present
authors \cite{brevik79,brevik70}.

In our opinion there is no strong connection between the Feigel
effect and relativity. The force on the fluid, or the momentum
transferred to it, are calculated assuming the fluid to be {\it at
rest}. Relativity is as little involved here as it is involved in
the description of the Walker-Lahoz experiment. An exceptional
case is, however,  if the Euler-Heisenberg Lagrangian is drawn
into consideration as a model to describe the ME effect (cf. for
instance, van Tiggelen {\it et al.} \cite{tiggelen06}).

9.  It is of interest to have an idea about the magnitude of the
effect that we have considered. Magnetoelectric birefringence is
actually found even in a vacuum, when there are strong crossed
external fields $\bf E_0$ and $\bf H_0$ present. The effect is
however extremely small. Let $\Delta n =n_B-n_E$ denote the
difference in the refractive index between the magnetic and
electric directions. Even with a strong magnetic field of 30 T and
an electric field of $10^8$ V/m the birefringence is only $\Delta
n \approx 8\times 10^{-23}$ \cite{rikken00}.

A more promising case is when one applies strong orthogonal fields
to a linear isotropic liquid. Thus Roth and Rikken \cite{roth02}
performed an experiment in which molecular liquids were placed in
such a strong field region. By passing laser light through the
liquid, perpendicular to the fields, they obtained a linear
relationship between the field strength and the MR birefringence.
With a magnetic field strength up to 17 T and an electric field of
2.5$\times 10^5$ V/m the ME birefringence was found to be of order
$\Delta n \sim 10^{-11}$. Thus  the ME effect is much larger in a
liquid than in a vacuum.

Naturally occurring anisotropies, the case that we have been
considering,  seem actually to be stronger.
 Thus the crystal FeGa$\rm O_3$ is known to be magnetoelectrically
 active with ME coefficients about $3\times 10^{-4}$ at low frequencies.
 In this crystal, as well as in analogous crystals like
  FeAl$\rm O_3$, anisotropies of order
 $10^{-4}$ are expected over a wide frequency range from DC to
 X-rays \cite{tiggelen06a}.


\begin{thebibliography}{99}
\bibitem{fotnote2}
Note, however, that Eq.~(8) in Ref.~\cite{ellingsen07} contains an
error: the factor $\mu$ in the denominator on the right hand side
should be  omitted.
\end{thebibliography}
\end{document}